\documentclass[12pt]{article}

\def\bracket#1#2{\bigl<#1\!\bigm|\!#2\bigr>}
\def\com#1#2{\bigl[#1,\,#2\bigr]}
\def\anticom#1#2{\bigl\{#1,\,#2\bigr\}}

\usepackage{amsmath,amssymb}

\newcommand{\real}{\mathbb{R}}
\newcommand{\comp}{\mathbb{C}}
\newcommand{\cS}{\mathcal{S}} 
\newcommand{\Piv}{\varPi}
\newcommand{\Thetav}{\varTheta}

\begin{document}

\title{\bf Operator Methods of \\
the Parabolic Potential Barrier}

\author{Toshiki Shimbori \\
{\footnotesize\it Institute of Physics, University of Tsukuba}\\
{\footnotesize\it Ibaraki 305-8571, Japan}}

\date{}

\maketitle

\begin{abstract}
 The one-dimensional parabolic potential barrier 
 dealt with in an earlier paper is re-examined 
 from the point of view of operator methods, 
 for the purpose of getting generalized Fock spaces. 
\end{abstract}

\thispagestyle{empty}

\setcounter{page}{0}

\pagebreak

 It is well known that the harmonic oscillator can be studied 
 by a operator method~\cite{dirac,jjs}. 
 In this paper we shall show that 
 the eigenvalue problem of the parabolic potential barrier~\cite{sk} 
 can be solved by a similar method. 
 The generalized eigenstates of this unstable system will form 
 {\it generalized Fock spaces}. 
 
 The Hamiltonian of the one-dimensional parabolic potential barrier is 
 \begin{equation}
  \hat{H}=\frac{1}{2m}\hat{p}^2 -\frac{1}{2} m\gamma^2 \hat{x}^2, 
   \label{2.1}
 \end{equation}
 where $m>0$ is the mass and $\gamma>0$ is 
 proportional to the square root of the curvature at the origin $x=0$. 
 The canonical coordinate and momentum $\hat{x}$ and $\hat{p}$ satisfy 
 the commutation relations 
 \begin{equation}
  \left.
   \begin{gathered} 
    {\com{\hat{x}}{\hat{p}}\equiv\hat{x}\hat{p}-\hat{p}\hat{x}
    =i\hslash}, \\
    \com{\hat{x}}{\hat{x}}=0,\,\,\,\com{\hat{p}}{\hat{p}}=0. 
   \end{gathered} \right\} \label{2.2} 
 \end{equation}
 Written in terms of the Schr\"{o}dinger or 
 the coordinate representatives, 
 the canonical coordinate and momentum give 
 \begin{equation}
  \left.
   \begin{aligned}
    \hat{x}&=x, \\
    \hat{p}&=-i\hslash\, d/dx, 
   \end{aligned} \right\} \label{2.3}
 \end{equation} 
 and the Hamiltonian \eqref{2.1} becomes
 \begin{equation}
  \hat{H} =-\frac{\hslash^2}{2m} \frac{d^2}{dx^2}
   -\frac{1}{2} m\gamma^2 x^2. \label{2.4} 
 \end{equation}
 
 We now introduce the {\it normal coordinates} 
 \begin{gather}
  \hat{b}^\pm\equiv\sqrt{\frac{m\gamma}{2\hslash}}
  \left(\hat{x}\pm\frac{1}{m\gamma}\hat{p}\right) 
  =\sqrt{\frac{m\gamma}{2\hslash}} 
  \left( x\mp\frac{i\hslash}{m\gamma} \frac{d}{dx}\right) 
  =\frac{1}{\sqrt{2}}\left( \xi\mp i\frac{d}{d\xi}\right), \label{2.5}\\ 
  \intertext{where} 
  \xi\equiv\beta x, \,\,\, 
  \beta\equiv\sqrt{\frac{m\gamma}{\hslash}}. \label{2.6} 
 \end{gather}
 Note that \eqref{2.5} are defined except for 
 an arbitrary phase factor. 
 These operators $\hat{b}^\pm$ are {\it essentially self-adjoint} on 
 a Schwartz space $\cS(\real)$. 
 (Cf. the normal coordinates $\hat{a}$ and $\hat{a}^\dag$, 
 being {\it adjoint operators}, for the harmonic oscillator.) 
 Further, two conditions are satisfied. 
 \begin{enumerate}
  \item $\cS(\real)$ is an invariant subspace of $\hat{b}^\pm$. 
  \item $\hat{b}^\pm$ is continuous on $\cS(\real)$. 
 \end{enumerate}  
 From the commutation relations \eqref{2.2} we obtain 
 \begin{equation}
 \left.
  \begin{gathered} 
   {\com{\hat{b}^+}{\hat{b}^-}=-i}, \\ 
   \com{\hat{b}^+}{\hat{b}^+}=0,\,\,\,\com{\hat{b}^-}{\hat{b}^-}=0. 
  \end{gathered} \right\} \label{2.7}
 \end{equation}
 The first of equations \eqref{2.7} gives the commutation relation 
 connecting $\hat{b}^+$ and $\hat{b}^-$. 
 (Cf. the commutation relation $\com{\hat{a}}{\hat{a}^\dag}=1$ 
 for the harmonic oscillator.) 
 One can express $\hat{H}$ in terms of $\hat{b}^+$ and $\hat{b}^-$ 
 and one finds 
 \begin{gather}
  \hat{H}=-\hslash\gamma\hat{N}, \label{2.8} 
  \intertext{where}
  \hat{N}\equiv\frac{1}{2}\anticom{\hat{b}^+}{\hat{b}^-}
  \equiv\frac{1}{2}
  \bigl(\hat{b}^+\hat{b}^- +\hat{b}^-\hat{b}^+\bigr). \label{2.9} 
 \end{gather}
 From \eqref{2.7} 
 \begin{align}
  \com{\hat{N}}{\hat{b}^\pm}&=\pm i\hat{b}^\pm \label{2.10} 
  \intertext{or}
  \com{\hat{H}}{\hat{b}^\pm}&=\mp i\hslash\gamma\hat{b}^\pm. \label{2.11}
 \end{align}
 Also, \eqref{2.7} lead to
 \begin{equation}
  \com{\hat{b}^\mp}{\bigl(\hat{b}^\pm\bigr)^n} 
   =\pm in \bigl(\hat{b}^\pm\bigr)^{n-1} \label{2.19}
 \end{equation}
 for any positive integer $n$. 
 
 We shall now work out the eigenvalue problem of $\hat{H}$. 
 Let us assume that there are {\it standard states} $u^\pm_0$ 
 satisfying 
 \begin{equation}
  \hat{b}^\mp u^\pm_0 =0. \label{2.12}
 \end{equation}
 If these equations are expressed in terms of representatives, 
 they give us
 \begin{equation}
  \left(\frac{d}{dx}\mp i\frac{m\gamma}{\hslash}x\right)
   u^\pm_0(x)=0 \label{2.13}
 \end{equation}
 with the help of \eqref{2.5}. The solutions of 
 these differential equations are 
 \begin{equation}
  u^\pm_0(x)=B^\pm_0 e^{\pm im\gamma x^2/2\hslash}, \label{2.14}
 \end{equation}
 where $B^\pm_0\in\comp$ are the numerical coefficients. 
 These solutions $u^\pm_0$ do not belong to a Lebesgue space $L^2(\real)$. 
 But they are {\it generalized functions} 
 in the conjugate space ${\cS(\real)}^\times$ of 
 the following Gel'fand triplet~\cite{sk,bogolubov,bohm}, 
 \begin{equation}
  \cS(\real)\subset L^2(\real)\subset{\cS(\real)}^\times. 
   \label{2.15}
 \end{equation}
 
 Let us treat the extensions $\bigl(\hat{b}^\pm\bigr)^\times$ 
 of the normal coordinates 
 to the conjugate space ${\cS(\real)}^\times$. 
 We should be able to apply them to a generalized function 
 $u\in{\cS(\real)}^\times$, 
 the products $\bigl(\hat{b}^\pm\bigr)^\times u$ 
 being defined by~\cite{bogolubov,bohm}
 $$\bracket{v}{\bigl(\hat{b}^\pm\bigr)^\times u}
 =\bracket{\hat{b}^\pm v}{u}$$
 for all functions $v\in\cS(\real)$. 
 Taking the representatives, we get
 $$\int_{-\infty}^{\infty}v(x)^* 
 \bigl[\bigl(\hat{b}^\pm\bigr)^\times u\bigr](x)dx=
 \int_{-\infty}^{\infty}\bigl[\hat{b}^\pm v\bigr](x)^* u(x)dx. $$
 We can transform the right-hand sides by partial integration and get
 $$\int_{-\infty}^{\infty}v(x)^* 
 \bigl[\bigl(\hat{b}^\pm\bigr)^\times u\bigr](x)dx=
 \int_{-\infty}^{\infty}v(x)^* \bigl[\hat{b}^\pm u\bigr](x)dx, $$ 
 since $v$ is a rapidly decreasing function and then 
 the contributions from the limits of integration vanish. 
 These give 
 $$\bracket{v}{\bigl(\hat{b}^\pm\bigr)^\times u}
 =\bracket{v}{\hat{b}^\pm u}, $$ 
 showing that 
 $$\bigl(\hat{b}^\pm\bigr)^\times u=\hat{b}^\pm u. $$
 Thus $\bigl(\hat{b}^\pm\bigr)^\times$ operating to a generalized function 
 have the meaning of $\hat{b}^\pm$ operating. 
 Similarly~\cite{sk}, 
 \begin{align*}
  \hat{N}^\times u &=\hat{N} u  
  \intertext{or}
  \hat{H}^\times u &=\hat{H} u. 
 \end{align*}
 
 Let us examine physical properties of the standard states. 
 The result of the operator $\hat{N}$ applied to 
 standard states $u^\pm_0$ is 
 \begin{equation}
  \hat{N}u^\pm_0 =\frac{1}{2}\hat{b}^\mp\hat{b}^\pm u^\pm_0 
   =\pm\frac{i}{2}u^\pm_0, \label{2.16} 
 \end{equation}
 with the help of \eqref{2.12} and \eqref{2.7}. From \eqref{2.8} 
 \begin{equation}
  \hat{H}u^\pm_0 =\mp\frac{i}{2}\hslash\gamma u^\pm_0. \label{2.17}
 \end{equation}
 Thus $u^\pm_0$ are generalized eigenstates 
 of $\hat{H}$ belonging to 
 the {\it complex energy eigenvalues} $\mp i\hslash\gamma/2$. 
 
 We can form 
 the generalized Fock spaces of the parabolic potential barrier 
 on the same lines as the harmonic oscillator~\cite{dirac,jjs}. 
 We now consider the following states: 
 \begin{equation}
  \bigl(\hat{b}^\pm\bigr)^n u^\pm_0. \label{2.18}
 \end{equation}
 The result of the operator $\hat{N}$ applied to these states is 
 \begin{equation}
  \hat{N}\bigl(\hat{b}^\pm\bigr)^n u^\pm_0 
   =\pm i\left( n+\frac{1}{2}\right)
   \bigl(\hat{b}^\pm\bigr)^n u^\pm_0, \label{2.20} 
 \end{equation}
 with the help of \eqref{2.19} and \eqref{2.12}. 
 Here we introduce the $n$th quantum states $u^\pm_n$, 
 being a numerical multiple of \eqref{2.18}, 
 which are satisfied by 
 \begin{gather}
  \hat{H}u^\pm_n = E^\pm_n u^\pm_n, \label{2.21} 
  \intertext{where} 
  E^\pm_n\equiv\mp i\left(n+\frac{1}{2}\right)\hslash\gamma \,\,\, 
  \left(n=0, 1, 2, \cdots\right). \label{2.22}
 \end{gather}
 Thus the states \eqref{2.18} are generalized eigenstates 
 of $\hat{H}$ belonging to 
 the {\it complex energy eigenvalues} $E^\pm_n$. 
 The representatives of 
 the $n$th quantum states can be obtained from \eqref{2.14}. 
 For any $f\in {\cS(\real)}^\times$ we find 
 $$\left(\mp i\frac{d}{d\xi}+\xi\right) f(\xi) 
 =e^{\mp i\xi^2/2}\left(\mp i\frac{d}{d\xi}\right)
 e^{\pm i\xi^2/2} f(\xi). $$
 Thus \eqref{2.18} give 
 \begin{align}
  \left[\frac{1}{\sqrt{2}}\left(\mp i\frac{d}{d\xi}+\xi\right)\right]^n 
  u^\pm_0(\xi) 
  &=\left(\frac{1}{\sqrt{2}}\right)^n e^{\mp i\xi^2/2} 
  \left(\mp i\frac{d}{d\xi}\right)^n e^{\pm i\xi^2/2} 
  u^\pm_0(\xi) \notag\\ 
  &=B^\pm_0\left(\frac{\mp i}{\sqrt{2}}\right)^n 
  e^{\pm i\xi^2/2} e^{\mp i\xi^2} 
  \frac{d^n}{d\xi^n} e^{\pm i\xi^2}. \label{2.23} 
 \end{align}
 Now define the polynomials $H^\pm_n(\xi)$ by~\cite{sk}
 \begin{equation}
  H^\pm_n(\xi)=\left(\mp i\right)^n e^{\mp i\xi^2} 
   \frac{d^n}{d\xi^n}e^{\pm i\xi^2}. \label{2.24}
 \end{equation}
 Inserting these expressions in \eqref{2.23}, 
 we get the representatives of the $n$th quantum states 
 \begin{equation}
  u^\pm_n(x)=B^\pm_n e^{\pm i\beta^2x^2/2} H^\pm_n(\beta x), 
   \label{2.25}
 \end{equation}
 where $B^\pm_n\in\comp$ are the numerical coefficients. 
 These numerical coefficients cannot be determined by 
 the normalizing condition~\cite{sk}. 
 
 Let us now see the properties of $n$th quantum states 
 under a parity or a space inversion. 
 The parity operator $\hat{\Piv}$ is a unitary operator on $\cS(\real)$ 
 and satisfies~\cite{jjs} 
 \begin{align*}
  \hat{\Piv}\hat{x}\hat{\Piv}^{-1}&=-\hat{x}, \\
  \hat{\Piv}\hat{p}\hat{\Piv}^{-1}&=-\hat{p}. 
 \end{align*}
 The parity operator applied to 
 the normal coordinates $\hat{b}^\pm$ is then 
 $$\hat{\Piv}\hat{b}^\pm\hat{\Piv}^{-1} =-\hat{b}^\pm $$ 
 from \eqref{2.5}. 
 The result of the parity operator applied to 
 the conditions \eqref{2.12} is 
 $$\hat{\Piv}\hat{b}^\mp u^\pm_0 
 =\hat{\Piv}\hat{b}^\mp\hat{\Piv}^{-1}\hat{\Piv}u^\pm_0 
 =-\hat{b}^\mp\hat{\Piv}u^\pm_0 = 0, $$
 showing that $\hat{\Piv}u^\pm_0$ are also standard states, i.e. 
 $$\hat{\Piv} u^\pm_0 = u^\pm_0. $$ 
 The phase factors are chosen unity, 
 because the representatives \eqref{2.14} are symmetrical of $x$. 
 We have further 
 $$\hat{\Piv}\bigl(\hat{b}^\pm\bigr)^n u^\pm_0 
 =\hat{\Piv}\hat{b}^\pm\hat{\Piv}^{-1}\cdots
 \hat{\Piv}\hat{b}^\pm\hat{\Piv}^{-1}\hat{\Piv} u^\pm_0 
 =(-)^n\bigl(\hat{b}^\pm\bigr)^n u^\pm_0, $$
 showing that 
 $$\hat{\Piv}u^\pm_n =(-)^n u^\pm_n. $$
 These equations assert that 
 the $n$th quantum states are eigenstates of the parity. 
 
 We shall now verify a time reversal. 
 The time-reversal operator $\hat{\Thetav}$ is 
 an antiunitary operator on $\cS(\real)$ and satisfies~\cite{jjs} 
 \begin{align*}
  \hat{\Thetav}\hat{x}\hat{\Thetav}^{-1}&=\hat{x}, \\
  \hat{\Thetav}\hat{p}\hat{\Thetav}^{-1}&=-\hat{p}. 
 \end{align*}
 The time-reversal operator applied to 
 the normal coordinates $\hat{b}^\pm$ is then 
 $$\hat{\Thetav}\hat{b}^\pm\hat{\Thetav}^{-1} =\hat{b}^\mp $$ 
 from \eqref{2.5}. 
 The result of the time-reversal operator applied to 
 the conditions \eqref{2.12} is 
 $$\hat{\Thetav}\hat{b}^\mp u^\pm_0 
 =\hat{\Thetav}\hat{b}^\mp\hat{\Thetav}^{-1}\hat{\Thetav}u^\pm_0 
 =\hat{b}^\pm\hat{\Thetav}u^\pm_0 = 0, $$
 showing that $\hat{\Thetav}u^\pm_0$ are also standard states, i.e. 
 $$\hat{\Thetav} u^\pm_0 = u^\mp_0. $$ 
 These time-reversed states contain arbitrary phase factors. 
 Proceeding as before, we have 
 $$\hat{\Thetav}\bigl(\hat{b}^\pm\bigr)^n u^\pm_0 
 =\hat{\Thetav}\hat{b}^\pm\hat{\Thetav}^{-1}\cdots
 \hat{\Thetav}\hat{b}^\pm\hat{\Thetav}^{-1}\hat{\Thetav} u^\pm_0 
 =\bigl(\hat{b}^\mp\bigr)^n u^\mp_0, $$
 showing that 
 $$\hat{\Thetav}u^\pm_n = u^\mp_n. $$
 We see in this way that a time reversal occurs 
 resulting in the interchange of the $n$th quantum states 
 $u^+_n$ and $u^-_n$. 
 
 Our work so far has been concerned with one instant of time. 
 We shall study finally the time evolution of 
 the parabolic potential barrier in the Heisenberg picture. 
 The Heisenberg equations of motion are 
 \begin{equation}
  \frac{d}{dt}\hat{b}^\pm(t) 
   =\frac{1}{i\hslash}\com{\hat{b}^\pm(t)}{\hat{H}}. \label{3.1} 
 \end{equation}
 With the help of \eqref{2.11}, these give 
 \begin{equation}
   \frac{d}{dt}\hat{b}^\pm(t)=\pm\gamma\hat{b}^\pm(t). \label{3.2}
 \end{equation}
 These equations can be integrated to give 
 \begin{equation}
  \hat{b}^\pm(t)=\hat{b}^\pm e^{\pm\gamma t}, \label{3.3}
 \end{equation}
 where $\hat{b}^\pm$ are normal coordinates \eqref{2.5}, 
 and are equal to the values of $\hat{b}^\pm(t)$ at time $t=0$. 
 The above solutions show that 
 $\hat{N}$ or $\hat{H}$ is also constant in the Heisenberg picture. 
 The canonical coordinate and momentum 
 in the Heisenberg picture are, from \eqref{2.5} and \eqref{3.3}, 
 \begin{equation}
  \left.
   \begin{aligned}
    \hat{x}(t)
    &=\hat{x}\cosh\gamma t +\hat{p}\sinh\gamma t/m\gamma, \\
    \hat{p}(t)
    &=\hat{x}m\gamma\sinh\gamma t +\hat{p}\cosh\gamma t 
    =m\Dot{\Hat{x}}(t). 
   \end{aligned} \right\} \label{3.7}
 \end{equation} 
 Equations \eqref{3.7} correspond to the hyperbolic orbits 
 in the classical theory. 
 
 The foregoing work provides two generalized Fock spaces 
 which are spanned by the tensor products of 
 $\left\{u^+_n\right\}_{n=0}^{\infty}$ and 
 $\left\{u^-_n\right\}_{n=0}^{\infty}$, respectively. 
 {\it The normal coordinates $\hat{b}^\pm$ are operators 
 of creation or annihilation 
 of a quantum of width $\hslash\gamma$}. 
 This interpretation has been given 
 by Takahashi in his book~\cite{ashi}. 
 For $\left\{u^\pm_n\right\}_{n=0}^{\infty}$, 
 the creation operators are $\hat{b}^\pm$ and 
 the annihilation operators are $\hat{b}^\mp$, 
 i.e. 
 the creation operator for $\left\{u^+_n\right\}_{n=0}^{\infty}$ are 
 the annihilation operator for $\left\{u^-_n\right\}_{n=0}^{\infty}$, 
 and vice versa. 
 These operators vary with time 
 according to the $e^{\pm\gamma t}$ law in the Heisenberg picture. 
 Thus the parabolic potential barrier 
 as a model of an unstable system will form a corner-stone 
 in the quantum theory of decay. 
 
 We may introduce 
 the essentially self-adjoint operators $\hat{d}^+$, $\hat{d}^-$ 
 satisfying the anticommutation relations 
 \begin{equation}
  \left.
   \begin{gathered}
    \anticom{\hat{d}^+}{\hat{d}^-}=1,\\
    \anticom{\hat{d}^+}{\hat{d}^+}=0,\,\,\,
    \anticom{\hat{d}^-}{\hat{d}^-}=0. 
   \end{gathered} \right\} \tag{\ref{2.7}$'$}\label{4.1}
 \end{equation}
 These relations are like \eqref{2.7} with an anticommutator 
 instead of a commutator. 
 Put
 \begin{equation}
  \hat{N}\equiv\frac{i}{2}
   \com{\hat{d}^+}{\hat{d}^-}, 
   \tag{\ref{2.9}$'$}\label{4.9} 
 \end{equation}
 the same as \eqref{2.9}. 
 We can find, by the above-mentioned method, 
 that the generalized eigenstates of $\hat{N}$ are 
 only the two alternatives of a twofold degenerate state 
 belonging to the complex eigenvalue $i/2$ or $-i/2$. 
 
 \section*{Acknowledgements}
 
 I would like to express my thanks to T. Kobayashi 
 for many valuable suggestions in the writing of this paper.

\end{document}